\documentclass[runningheads]{llncs}
\usepackage{times}
\usepackage{latexsym}
\usepackage{amsfonts}
\usepackage{graphicx}
\usepackage{url}
\usepackage{array}
\usepackage{pgfplots}
\usepgfplotslibrary{groupplots}
\pgfplotsset{compat=newest}

\title{Document Network Projection in Pretrained Word Embedding Space}
\author{Antoine Gourru\inst{1}\orcidID{0000-0003-3571-2430} 
\and Adrien Guille\inst{1}\orcidID{0000-0002-1274-6040} 
  \and Julien Velcin\inst{1}\orcidID{0000-0002-2262-045X}
   \and Julien Jacques\inst{1}\orcidID{0000-0003-4808-2781}
  }
  \authorrunning{A. Gourru et al.}
\institute{ Universit\'e de Lyon, Lyon 2 \\
  ERIC EA3083\\\email{\{firstname\}.\{lastname\}@univ-lyon2.fr}}

\begin{document}
\bibliographystyle{splncs04}
\maketitle

\begin{abstract}

We present Regularized Linear Embedding (RLE), a novel method that projects a collection of linked documents (e.g. citation network) into a pretrained word embedding space. In addition to the textual content, we leverage a matrix of pairwise similarities providing complementary information (e.g., the network proximity of two documents in a citation graph). We first build a simple word vector average for each document, and we use the similarities to alter this average representation. The document representations can help to solve many information retrieval tasks, such as recommendation, classification and clustering. We demonstrate that our approach outperforms or matches existing document network embedding methods on node classification and link prediction tasks. Furthermore, we show that it helps identifying relevant keywords to describe document classes. 
\keywords{Document Network Embedding \and Representation Learning}
\end{abstract}

\section{Introduction}

Information retrieval methods require relevant compact vector space representations of documents.
The classical bag of words cannot capture all the useful semantic information.
Representation Learning is a way to go beyond and boost the performances we can expect in many information retrieval tasks \cite{devlin2018bert}. It aims at finding low dimensional and dense representations of high dimensional data such as words \cite{mikolov2013distributed} and documents \cite{le2014distributed,arora2016simple}. In this latent space, proximity reflects semantic closeness. Many recent methods use those representations for information retrieval tasks:  capturing user interest \cite{seyler2018information}, query expansion \cite{kuzi2016query}, link prediction and document classification \cite{yang2015network}. 

In addition to the textual information, many corpora include links between documents, such as bibliographic networks (e.g., scientific articles linked with citations or co-authorship) and social networks (e.g., tweets with ReTweet relations). This information can be used to improve the accuracy of document representations. Several recent methods \cite{yang2015network,liu2018content} study the embedding of networks with textual attributes associated to the nodes. Most of them learn continuous representations for nodes independently of a word-vector representation. That is to say, documents and words do not \textit{lie} in the same space. It is interesting to find a common space to represent documents and words when considering many tasks in information retrieval (query expansion) and document analysis (description of document clusters). Our approach allows to represent documents and words in the same semantic space. The method can be applied with word embedding learned on the data  with any state-of-the art method \cite{devlin2018bert,mikolov2013distributed}, or with embeddings that were previously learned\footnote{e.g., \url{https://fasttext.cc/}} to reduce the computation cost. Contrary to many existing methods that make use of deep and complex neural networks (see Section 2 for related works), our method is fast, and it has only one parameter to tune.

\begin{figure}
    \centering
    \includegraphics[width=0.3\textwidth]{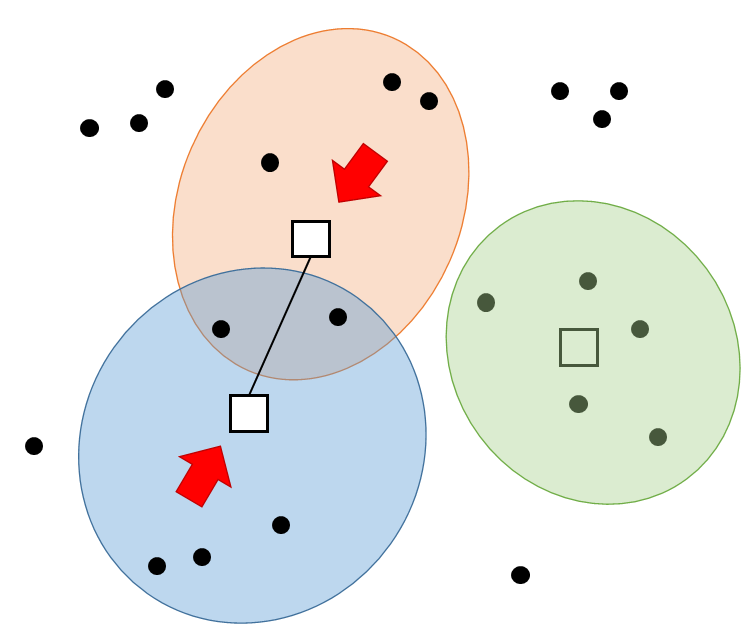}
    \caption{Our method performs smoothing (represented as red arrows) on the documents' centroid representations (the square blocks). As the document in the blue circle (dots are words) is connected to the orange one, their representations get closer. The document in the green circle is isolated, thus it remains unchanged by the smoothing effect.}
    \label{fig:illu}
\end{figure}

We propose to construct a weight vector for each document using both textual and network information. We can then project the documents into the prelearned word vector space using this vector (see Figure~\ref{fig:illu}). The method is straightforward to apply, as it only requires applying well studied word embedding methods and matrix multiplication. We show in Section 4 that it outperforms or matches existing methods in classification and link prediction tasks and we demonstrate that projecting the documents into the word embedding space can provide semantic insights.

\vspace{-1cm}
\section{Related Work}

Several methods study the embedding of paragraph or short documents such as \cite{le2014distributed}, generalizing the seminal word2vec models proposed by \cite{mikolov2013distributed}. These approaches go beyond the simple method that consists in building a weighted average of representations of words that compose the document. For example in \cite{arora2016simple}, authors propose to perturb weights for word average projection using Singular Value Decomposition (SVD). This last approach inspired our work as they show that word average is often a relevant baseline that can be improved in some cases using contextual smoothing.

As stated above, many corpora are structured in networks, providing additional information on documents semantics. TADW \cite{yang2015network} is the first method that deals with this kind of data. It formulates network embedding \cite{perozzi2014DeepWalk} as a matrix tri-factorization problem to integrate textual information. Subsequent methods mainly adopt neural network based models: STNE \cite{liu2018content} extends the seq2seq models, Graph2Gauss \cite{bojchevski2018g2g} learns both representations and variances via energy based learning, and VGAE \cite{kipf2016vgae} adopts a variational encoder. Even if these approaches yield good results, they require tuning a lot of hyperparameters.
Two methods are based on factorization approaches: GVNR-t \cite{brochier2019gvnr}, that extends GloVe \cite{pennington2014glove}, and AANE \cite{huang2017aane}. None of these methods learn documents and words embedding in the same space. In \cite{le2014distributed} and \cite{ailem2017non}, authors represent them in a comparable space.  Yet, they do not consider network information, as opposed to LDE \cite{wang2016linked}. Nonetheless, this last method requires labels associated with nodes, making it a supervised approach. Our method projects the documents and the words into the same space in an unsupervised fashion, with only one hyperparameter to tune. We will now present the formulation of this approach.

\section{RLE: Document projection with smoothing}

In this section, we present our model to build vector representations for a collection of linked documents. From now on, we will refer to our method as Regularized Linear Embedding (RLE). Matrices are in capital letters, and if $X$ is a matrix, we write $x_i$ the $i$-th row of $X$. From a network of $n$ nodes, we extract a pairwise similarity matrix $S \in \mathbb{R}^{n \times n}$, computed as $S = \frac{A + A^2}{2}$ with $A$ the transition matrix of the graph.
Similarly to \cite{yang2015network}, this matrix considers both first and second order similarities. $v$ is the number of words in the vocabulary. The corpus is represented as a document-term matrix $T \in \mathbb{R}^{n \times v}$, with each entry of $T$ being the relative frequency of a word in a given document.

With $U \in \mathbb{R}^{v \times k}$ a matrix of pretrained word embeddings in dimension $k$, our goal is to build a matrix $D \in \mathbb{R}^{n \times k}$ of document embeddings , in the same space as the word embeddings. We build, for each document, a weight vector $p_i \in \mathbb{R}^v$,  stacked in a matrix $P$ and define the embedding of a document as $d_{i} = p_iU$.  We construct $p_i$ as follows: we first compute a smoothing matrix $B \in \mathbb{R}^{n \times v}$ with:
\begin{equation}
    b_i = \frac{1}{\sum_j S_{i,j}}\sum_j S_{i,j}t_j.
\end{equation}
Each row $b_i$ of this matrix is a centroid of the initial document-term frequency matrix $T$, weighted by the similarity between the document $i$ and each of the other documents. 
Then, we compute the weight matrix P according to $T$ and $B$, in matrix notation:
\begin{equation}
    P = (1 - \lambda) T + \lambda B,
\end{equation}
where $\lambda\in\lbrack0,1\rbrack$ controls the smoothing intensity. Then, we compute $D = PU$. Our method implies matrix multiplication and normalization only, making it fast and easily scalable. When $\lambda = 0$, $P=T$, thus, we recover the word average method. When $\lambda = 1$, we obtain $P=B$ and thus embed the documents with respect to the contextual information only (i.e., the similar documents). We illustrate the effect of smoothing in Figure~\ref{fig:illu}.

\section{Experiments}

In this section, we present our experimental results on classification and link prediction tasks, followed by a qualitative analysis of document representations.

We use two citation networks: Cora \cite{tu2017cane} and DBLP \cite{tang2008arnetminer,pan16DBLP}. We also use New York Times articles (\url{https://www.nytimes.com/}) from January 2007. We create a link between pairs of articles sharing a common tag. The class corresponds to the article section. Cora contains 2,211 labeled documents (7 classes) with 5,001 citation links. The dataset includes the abstract of each article. the New York Times dataset (Nyt) contains 5,135 documents, 3,050,513 edges and 4 classes. Dblp has 60,744 documents (4 classes) and  52,914 edges. It includes the title of the articles only. After pruning the vocabulary (removing stop words, filtering word occurring in more than 25\% of the corpus and less than 10 times), we obtain vocabularies made of 2,421 features for the Cora dataset, 6,407 for the Nyt dataset, and 3,763 for Dblp.

All embeddings are in dimension 160. We use DeepWalk with 40 walks of length 40, and a window of size 10. We also experiment with Latent Semantic Analysis (LSA) \cite{deerwester1990indexing} and a concatenation of LSA and DeepWalk representations in dimension 80 as done by \cite{yang2015network}, referred as ``Concatenation''. We also compare the performance of RLE with recent methods that embed attributed networks: STNE, Graph2Gauss, GVNR-t, VGAE, AANE and TADW. For STNE, we set the depth to 1 which leads to
the best scores in our experiments. For Graph2Gauss, we set K=1, depth=1. We use default architecture for VGAE and determine optimal $\lambda$ and $\rho$ for AANE, and $x_{min}$ for GVNR-t. For TADW, we use LSA in dimension 200 as a textual feature matrix and set regularization to 0.2, following authors' recommendation. For each method, we use the implementation provided by the authors. We discard LDE since it is semi supervised and will not lead to a fair comparison. 
 
RLE needs prelearned word representations. Hence, we build word vectors using Skip-gram with negative sampling \cite{mikolov2013distributed}. We use the implementation in gensim\footnote{\url{https://radimrehurek.com/gensim/}}, with window size of 15 for Cora, 10 for Nyt and 5 for DBLP, and 5 negative examples for both. The procedure is fast (46 seconds for Cora, 84 on DBLP and 42 on Nyt). Similarly to baselines methods, we use the value of $\lambda$ (0.7) that produces the optimal results on both datasets (see Figure~\ref{fig:lambda}).

\subsection{Quantitative results}
\vspace{-0.5cm}
\begin{table*}
\centering
\caption{Comparison of Micro-F1 results on a classification task for different train/test ratios. The best score is in bold, second best is underlined. Execution time order is presented in seconds (Time).}\label{tab:resC}
\begin{tabular}{r|ccc|c||ccc|c|}
 & \multicolumn{4}{c||}{Cora} & \multicolumn{4}{c|}{Dblp}\\
train/test ratio & 10\% & 30\% & 50\% & Time& 10\% & 30\% & 50\% & Time\\
 \hline
DeepWalk & 70.6 (2.0) & 77.2 (0.9) & 81.0 (0.7) & $10^{1}$ & 52.3 (0.4) & 53.4 (0.1) & 53.5 (0.2) & $10^{2}$ \\
LSA & 72.3 (1.9) & 79.0 (0.7) & 80.6 (0.7) & $10^{-2}$ & 73.5 (0.2) & 74.1 (0.1) & 74.2 (0.2) & $10^{1}$ \\
Concatenation & 71.4 (2.1) & 80.5 (1.0) & 84.0 (1.1) & $10^{1}$ & \underline{77.5} (0.2) & \underline{78.0} (0.1) & \underline{78.2} (0.2) & $10^{2}$ \\\hline
TADW & 81.9 (0.8) & 86.3 (0.8) & \underline{87.4} (0.8) & $10^{-1}$ & 74.8 (0.1) & 75.3 (0.2) & 75.5 (0.1) & $10^{1}$ \\
AANE & 79.8 (0.9) & 83.3 (1.1) & 84.4 (0.7) & $10^{-1}$ & 73.3 (0.1) & 73.9 (0.1) & 74.2 (0.2) & $10^{2}$ \\
GVNR-t & \underline{83.7} (1.2) & \underline{86.4} (0.7) & 87.0 (0.8) & $10^{1}$ & 69.6 (0.1) & 70.1 (0.1) & 70.2 (0.2) & $10^{2}$ \\\hline
VGAE & 72.3 (1.7) & 79.2 (0.9) & 81.1 (0.7) & $10^{1}$ &\multicolumn{3}{c|}{Memory overflow}&-- \\
G2G & 79.0 (1.5) & 83.7 (0.8) & 84.8 (0.7) & $10^{1}$ & 70.8 (0.1) & 71.3 (0.2) & 71.5 (0.2) & $10^{2}$ \\
STNE & 79.4 (1.0) & 84.7 (0.7) & 86.7 (0.8) & $10^{2}$ & 73.8 (0.2) & 74.4 (0.1) & 74.5 (0.1) & $10^{4}$ \\
RLE & \textbf{84.0} (1.3) & \textbf{86.9} (0.5) & \textbf{87.7} (0.6) & $10^{1}$ & \textbf{79.8} (0.2) & \textbf{80.9} (0.2) & \textbf{81.2} (0.1)& $10^{1}$ \\
\end{tabular}
\begin{tabular}{r|ccc|c|}
 & \multicolumn{4}{c|}{Nyt} \\
 train/test ratio& 10\% & 30\% & 50\% & Time\\
 \hline
DeepWalk & 66.9 (0.7) & 68.2 (0.3) & 68.7 (0.9) & $10^{2}$ \\
LSA & 71.6 (1.0) & 75.7 (0.7) & 76.7 (0.7) & $10^{-2}$ \\
Concatenation  & \textbf{77.9} (0.3) & \textbf{80.0} (0.5) & \textbf{81.1} (0.7) & $10^{2}$ \\\hline
TADW & 75.8 (0.5) & 78.4 (0.5) & 79.4 (0.4) & $10^{1}$ \\
AANE & 71.7 (0.5) & 75.6 (0.8) & 76.9 (1.1) & $10^{1}$ \\
GVNR-t & 74.3 (0.4) & 76.0 (0.6) & 76.7 (0.6) & $10^{2}$ \\
\hline
VGAE & 68.1 (0.8) & 69.3 (0.9) & 70.1 (0.6) & $10^{2}$ \\
G2G & 69.0 (0.5) & 70.5 (0.7) & 71.5 (0.8) & $10^{2}$ \\
STNE & 75.1 (0.7) & 77.3 (0.5) & 78.1 (0.6) & $10^{2}$ \\
\hline
RLE & \underline{77.7} (0.7) & \underline{79.3} (0.5) & \underline{80.0} (0.6) & $10^{1}$ \\
\end{tabular}
\end{table*}

\begin{table}
\centering

\caption{Comparison of AUC results on a link prediction task for different percents of edges hidden. The best score is in bold, second best is underlined.}\label{tab:resLP}
\begin{tabular}{r|c|c||c|c|}
 & \multicolumn{2}{c||}{Cora} & \multicolumn{2}{c|}{Dblp}\\
\% edges hidden& 50\% &  25\% & 50\% &  25\% \\\hline
DeepWalk          & 73.2 (0.6) & 80.9 (1.0) & \textbf{89.7} (0.0) & \textbf{93.2} (0.2) \\
LSA               & 87.4 (0.6) & 87.2 (0.8) & 54.2 (0.1) & 54.8 (0.0)   \\
Concatenation & 77.9 (0.3) & 83.7 (0.8) & 88.8 (0.0) & \underline{92.6} (0.3) \\\hline
TADW              & 90.1 (0.4) & 93.3 (0.4) &  61.2 (0.1) &65.0 (0.5) \\
AANE              & 83.1 (0.8) & 86.6 (0.8) & 67.4 (0.1) &66.5 (0.1) \\
GVNR-t            & 83.9 (0.9) & 91.5 (1.1) & 88.1 (0.3) & 91.4 (0.1) \\
VGAE               & 87.1 (0.4) & 88.2 (0.7) &\multicolumn{2}{c|}{Does not scale} \\
Graph2Gauss       & \underline{92.0} (0.3) & \underline{93.8} (1.0) & 88.0 (0.1) &92.1 (0.5) \\
STNE              & 83.1 (0.5) & 90.0 (1.0) & 45.6 (0.0) & 53.4 (0.1) \\\hline
RLE               & \textbf{94.3} (0.2) & \textbf{94.8} (0.2) & \underline{89.3} (0.1) & 91.2 (0.2)        
\end{tabular}
\end{table}

\begin{figure}
\begin{tikzpicture}
\begin{groupplot}[group style={
                      group name=f1score,
                      group size= 3 by 1},width=0.35\textwidth,height=2.5cm]

    \nextgroupplot[title=\textbf{(a) Cora},  ytick = {80, 84, 88},xtick = {0, 0.7, 1}, xlabel=$\lambda$, ylabel=Micro F1 score]
        \addplot[mark=none, color=black, thick] table [x=lambda, y=CORA, col sep=comma] {csv/cora_lambda.csv};
    \nextgroupplot[xshift=0.3cm,title=\textbf{(b) Dblp}, ytick = {74, 78, 82},xtick = {0, 0.7,  1}, xlabel=$\lambda$, ylabel=Micro F1 score]
        \addplot[mark=none, color=black, thick] table [x=lambda, y=dblp, col sep=comma] {csv/dblp_lambda.csv};
    \nextgroupplot[xshift=0.3cm,title=\textbf{(a) Nyt}, xtick = {0, 0.7, 1}, xlabel=$\lambda$, ylabel=Micro F1 score]
        \addplot[mark=none, color=black, thick] table [x=lambda, y=NYT, col sep=comma] {csv/nyt_lambda.csv};
\end{groupplot}
\end{tikzpicture}
\caption{Impact of $\lambda$ on RLE in terms of document classification for $d=160$. Optimum is achieved around 0.7 on each dataset (Cora, Nyt: 0.7, Dblp: 0.65).}
\label{fig:lambda}
\end{figure}
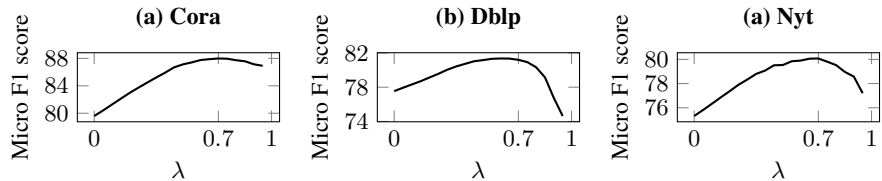

We evaluate RLE in its ability to separate documents by classes in the embedding space and to predict links between documents. We perform SVM with L2 regularization on the vector representations of documents and report Micro F1 scores for different train/test ratios in Table~\ref{tab:resC}. The regularisation strength is fixed through grid search. We also report computation times in second. For link prediction, we hide a percent of edges and compare the cosine similarity between hidden pairs and negative examples of unconnected documents. We report the Area Under the Roc Curve in Table~\ref{tab:resLP}.

In the classification task, RLE outperforms existing methods on Cora and Dblp, and is the second best method on Nyt. Interestingly, GVNR-t performs well with few training example, while TADW become second with 50\% of training examples. Let us highlight that RLE runs fast, it is even faster than AANE on Dblp. Additionnaly, it is up to four orders of magnitude faster than STNE on Dblp. Additionally Figure~\ref{fig:lambda} shows that the optimal lambda values are similar for both datasets. Its tuning is not that crucial since RLE outperforms the baselines with $\lambda \in \lbrack 0.6,0.85 \rbrack $ on Cora, $\lambda \in \lbrack 0.15,0.85 \rbrack$ on DBLP, and every methods except Concatenation for $\lambda \in \lbrack 0.45,0.8 \rbrack$ on Nyt .

In link prediction, RLE outperforms existing methods on Cora, while DeepWalk yields better results than baselines on Dblp. This might be due to the shortness of the documents (mean length is 6 while it is 49 for Cora): the textual information may not be as informative as the network information for link prediction.

\subsection{Qualitative insights}\label{sec:qualIns}

We compute a vector representation for a class by computing the centroid of the representations of the documents inside this class. We present the closest words to this representation in term of cosine similarity, which provides a general description of the class. In Table~\ref{tab:resQuali}, we present a description using this method for the first four classes of the Cora Dataset. We also provide most weighted terms when computing the mean of documents $tf \cdot idf$ of the class. The  $tf \cdot idf$ method produces too general words, such as ``learning'', ``algorithm'' and ``model''. RLE seems to provide specific words, which makes the descriptions more relevant.

\begin{table*}
\centering
\scriptsize
\caption{Classes Description with our method as opposed to $tf \cdot idf$. Words that are repeated across classes are in bold. RLE produces more discriminative descriptions}\label{tab:resQuali}
\begin{tabular}{cc|cc|cc|cc}
\multicolumn{8}{c}{\textbf{Cora}}\\\hline
\multicolumn{2}{c|}{Class 1} & \multicolumn{2}{c|}{Class 2}    & \multicolumn{2}{c|}{Class 3} & \multicolumn{2}{c}{Class 4}     \\
RLE        & $tf \cdot idf$ & RLE           & $tf \cdot idf$ & RLE        & $tf \cdot idf$ & RLE            & $tf \cdot idf$ \\\hline
hebbian    & neural         & reinforcement & \textbf{learning}       & posterior  & bayesian       & pac            & \textbf{learning}       \\
network    & network        & discounted    & reinforcement  & gibbs      & \textbf{model}          & learnability   & \textbf{algorithm}      \\
layers     & networks       & qlearning     & control        & bayesian   & models         & polynomialtime & algorithms     \\
multilayer & \textbf{learning}   & rl            & state          & mcmc       & \textbf{algorithm}      & dnf            & \textbf{model}          \\
filters    & \textbf{model}          & multiagent    & policy         & sampler    & belief         & queries        & decision   
\end{tabular}
\end{table*}
\vspace{-1cm}
\section{Conclusion}

In this article, we presented the RLE method for embedding documents that are organized in a network.
Despite its simplicity, RLE shows state-of-the art results for the three considered datasets. It is faster than most recent deep-learning methods. Furthermore, it provides informative qualitative insights. Future works will concentrate on automatically tuning $\lambda$, and exploring the effect of the similarity matrix $S$.

\bibliography{bibli}

\end{document}